\documentclass[11pt]{article}
\usepackage{moriond}

\usepackage{hyperref}

\def\MET{E\llap{/\kern1.5pt}_T}

\newcommand{\Photo}{}

\begin{document}
\vspace*{4cm}
\title{MULTILEPTON AND MULTIPHOTON SIGNATURES OF SUPERSYMMETRY AT THE LHC}

\author{ CHRISTOFFER PETERSSON}

\address{Physique Th\'eorique et Math\'ematique,\\
Universit\'e Libre de Bruxelles, C.P. 231, 1050 Brussels, Belgium.\\
International Solvay Institutes, Brussels, Belgium.\\
Department of Fundamental Physics,\\
Chalmers University of Technology, 412 96 G\"oteborg, Sweden.\\
$\mathrm{Email: christoffer.petersson@ulb.ac.be,~christoffer.petersson@chalmers.se}$
}

\maketitle\abstracts{ 
Motivated by the absence of any clear signal of physics beyond the Standard Model at the LHC after Run I, we discuss one possible slight hint of new physics and one non-minimal extension of the Standard Model. In the first part we provide a tentative explanation of a small excess of multilepton events, observed by the CMS collaboration, by means of a simplified model of gauge mediated supersymmetry breaking. In the second part we discuss how the standard phenomenology of gauge mediation can be significantly modified if one makes the non-minimal assumption that supersymmetry is broken in more than one hidden sector. Such multiple hidden sector models involve light neutral fermions called pseudo-goldstini and, due to the extra decay steps they induce, where soft photons are emitted, these models give rise to multiphoton plus missing energy signatures. We discuss why the existing LHC searches are poorly sensitive to these model and we propose new searches designed to probe them.}

\section{Introduction}
The discovery of a scalar boson, with properties in agreement with the predictions of the Standard Model (SM), reinforces the desire to understand the origin of the electroweak scale. The quadratic sensitivity of the scalar mass to ultraviolet physics suggests the presence of new states beyond the SM (BSM) at or below the TeV scale. But, in spite of the impressive range of searches performed by the LHC collaborations during Run I, no such new states have been observed so far. While these null results may be taken as an indication of the absence of new states in this energy range, they may also be taken as a motivation to push forward in new directions in the exploration of TeV scale physics.   

Supersymmetric (SUSY) extensions of the SM have the potential to both stabilize the electroweak scale and explain why it is hierarchically smaller than the Planck scale. The current bounds on superpartner masses are discomforting, but one should perhaps keep in mind that many searches are designed to probe particular SUSY extensions which are {\it minimal} in terms of their particle content and the underlying assumptions. However, given the non-minimality of, for example, the particle content of the SM, with three generations of quarks and leptons with a hierarchical mass spectrum, it could be that minimality is not a good guiding principle. 

By going beyond minimality in terms of model building, non-standard phenomenology can easily arise,  with new search channels opening up and/or with standard search channels closing down. In the fist part of this note, by allowing for spectra beyond those of minimal models, we discuss an example of a BSM model that both fits a slight excess in the data and that predicts non-standard signatures which are currently not being targeted at the LHC. In the second part, we discuss a scenario where a deviation from the minimal model building assumptions opens up new search channels, while evading constraints coming from the standard ones.            

We consider simplified models based on the framework of gauge mediated SUSY breaking (GMSB), with R-partity conservation. In Section \ref{multileptons}, based on the paper \cite{multileptons}, we do the exercise of explaining a slight excess in terms of multilepton events observed by the CMS collaboration \cite{CMSmultileptons}. We provide a simple model that can explain the excess, without being excluded by any other data, and we discuss how to best probe this model.  In Section \ref{multiphotons}, based on the paper \cite{multiphotons}, we discuss how the standard phenomenology of GMSB is modified if SUSY is broken in more than one hidden sector.\footnote{For different discussions and aspects of multiple hidden sector models in the context of gravity mediation and gauge mediation, see \cite{goldstini} and \cite{goldstiniGMSB}, respectively.} In such multiple hidden sector models, the final state spectrum is typically softer than in standard GMSB, which implies that existing LHC searches are not very sensitive to these kind of models. The upshot is that these models typically give rise to additional (soft) photons in the final state, and we propose new searches designed to probe them.  

\section{Multilepton signatures}
\label{multileptons}

Let us start by discussing the small excess observed by the CMS collaboration in a search for events with three or more leptons with 19.5 fb${}^{-1}$ of data at $\sqrt{s}=8$ TeV \cite{CMSmultileptons}. This small excess was seen in the final state category of events with three electrons or muons,\,\footnote{Out of these three electrons or muons, there is one opposite sign same flavor (OSSF) lepton pair whose invariant mass is outside a window of $\pm$15 GeV around the $Z$ boson mass.} one hadronically decaying tau lepton ($\tau_h$), low hadronic activity\,\footnote{The scalar sum of the jet $p_T$ values, denoted by $H_T$, is required to be below 200 GeV.} and no tagged b-jets. In this category, CMS observed (expected) \mbox{15 ($7.5\pm2.0$), 4 ($2.1\pm0.5$) and 3 ($0.60\pm0.24$) events} in the three bins of missing transverse energy $\MET{<}50$\,GeV, 50${<}\MET{<}100$\,GeV and $\MET{>}100$\,GeV, respectively. The  probability to observe 22 events in the combined $\MET$-range, when $10.1\pm2.4$ events were expected, is about 1\%. However, when taking into account the fact that they search in 64 independent categories, the probability for this fluctuation in the combined $\MET$-range is about 50\%, while the joint probability to observe such an excess in all the three $\MET$-bins is about 5\% \cite{CMSmultileptons}.

\begin{figure}
\centerline{\includegraphics[width=0.55\linewidth]{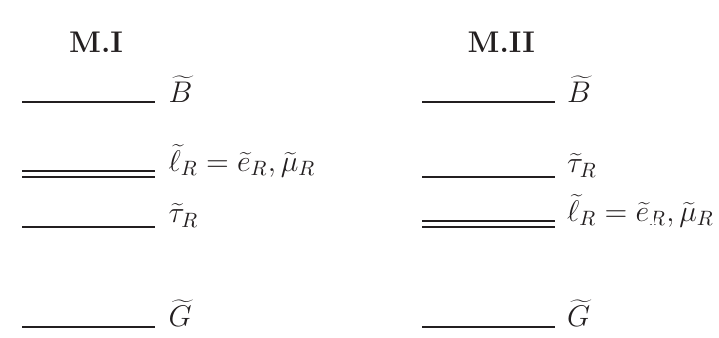}}
\caption{Spectra for the simplified models {\bf M.I} and {\bf M.II}.}
\label{fig:models}
\end{figure}

The most likely explanation for this slight excess is that it is due to a statistical fluctuation and that it will go away with more data. Nevertheless, we take the opportunity to perform the exercise of trying to fit this excess with some BSM physics. We consider two simplified models of GMSB, denoted by {\bf M.I} and {\bf M.II}, with spectra given in Figure \ref{fig:models}. These models were studied in \cite{CMSmultileptons}. Here we extend that study by taking into account the exclusion bounds arising from pair production of the next-to-lightest SUSY particle (NLSP), determining the best fit model, considering other relevant searches and  discussing prospects and possible new searches designed to probe the best fit model.  

Concerning the particle content of the models in Figure \ref{fig:models}, as always in GMSB, the lightest SUSY particle (LSP) is the nearly massless gravitino $\widetilde{G}$. In model {\bf M.I}/{\bf M.II} we take the NLSP to be the right-handed stau/sleptons, $\tilde{\tau}_R/\tilde{\ell}_R$, where ``slepton"  refers to either a selectron or a smuon, $\tilde{\ell}_R=\tilde{e}_R,\tilde{\mu}_R$. The next-to-NLSP (NNLSP) is the $\tilde{\ell}_R/\tilde{\tau}_R$, while the Bino $\widetilde{B}$ is taken to be heavier. All remaining superpartners are assumed to be sufficiently heavy and effectively decoupled. While such a decoupling is typically not possible in minimal GMSB models, where the relations among the soft masses are completely determined in terms of the gauge quantum numbers, it is possible to realize such spectra within the framework of \mbox{General Gauge Mediation \cite{GGM}.} See \cite{selectronNLSP} for a complete characterization of models realizing the non-standard GMSB spectrum of the simplified model {\bf M.II} in Figure \ref{fig:models}.   

Concerning the decay channels, since we assume R-parity, the NLSP only has one decay mode, i.e.~to its SM partner and the gravitino. In contrast, the NNLSP has two possible decay channels, either the two-body decay to its SM partner and the gravitino, or the three-body decay, via an off-shell Bino, to the NLSP. For the parameter space region we are interested in, where the gravitino mass is in the range 0.1\,eV${<} \,m_{3/2} {<}
\,10$\,eV, the NNLSP coupling to the gravitino is strongly suppressed compared to the gauge couplings entering the three-body decay, and the dominant NNLSP decay mode is the three-body decay.   

\begin{figure}
\centering
\vspace{.24cm}
 \includegraphics[width=.37\textwidth]{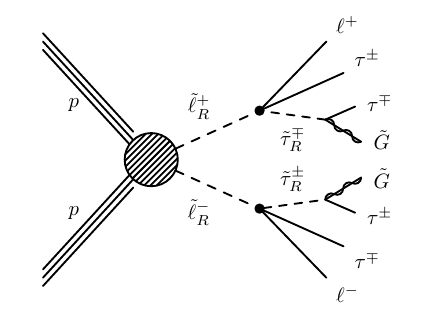}
 \hspace{.3cm}
 \includegraphics[width=.37\textwidth]{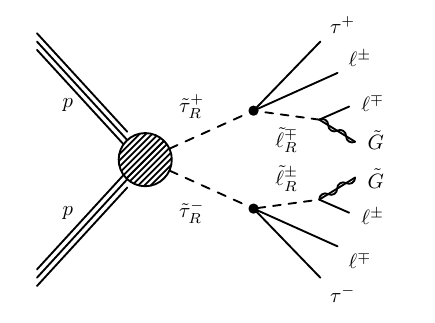}
 \caption{\label{fig:processes} The NNLSP pair production processes for models {\bf M.I} (left) and {\bf M.II} (right).}
\end{figure}

At the LHC, the models {\bf M.I} and {\bf M.II} give rise to the processes shown in Figure \ref{fig:processes}, and the final states $4\tau+2\ell+\MET$ and $2\tau+4\ell+\MET$, respectively. Hence, NNLSP pair production gives rise to multilepton events which could be relevant for the CMS search \cite{CMSmultileptons}. In order to see if we can fit the excess in \cite{CMSmultileptons}, we simulate the two processes in Figure \ref{fig:processes} at the LHC and analyze \mbox{19.5 fb${}^{-1}$} of data at $\sqrt{s}=8$ TeV, with kinematic and geometric cuts applied in accordance with the CMS search.  

\begin{figure}
\centering
 \includegraphics[width=.32\textwidth]{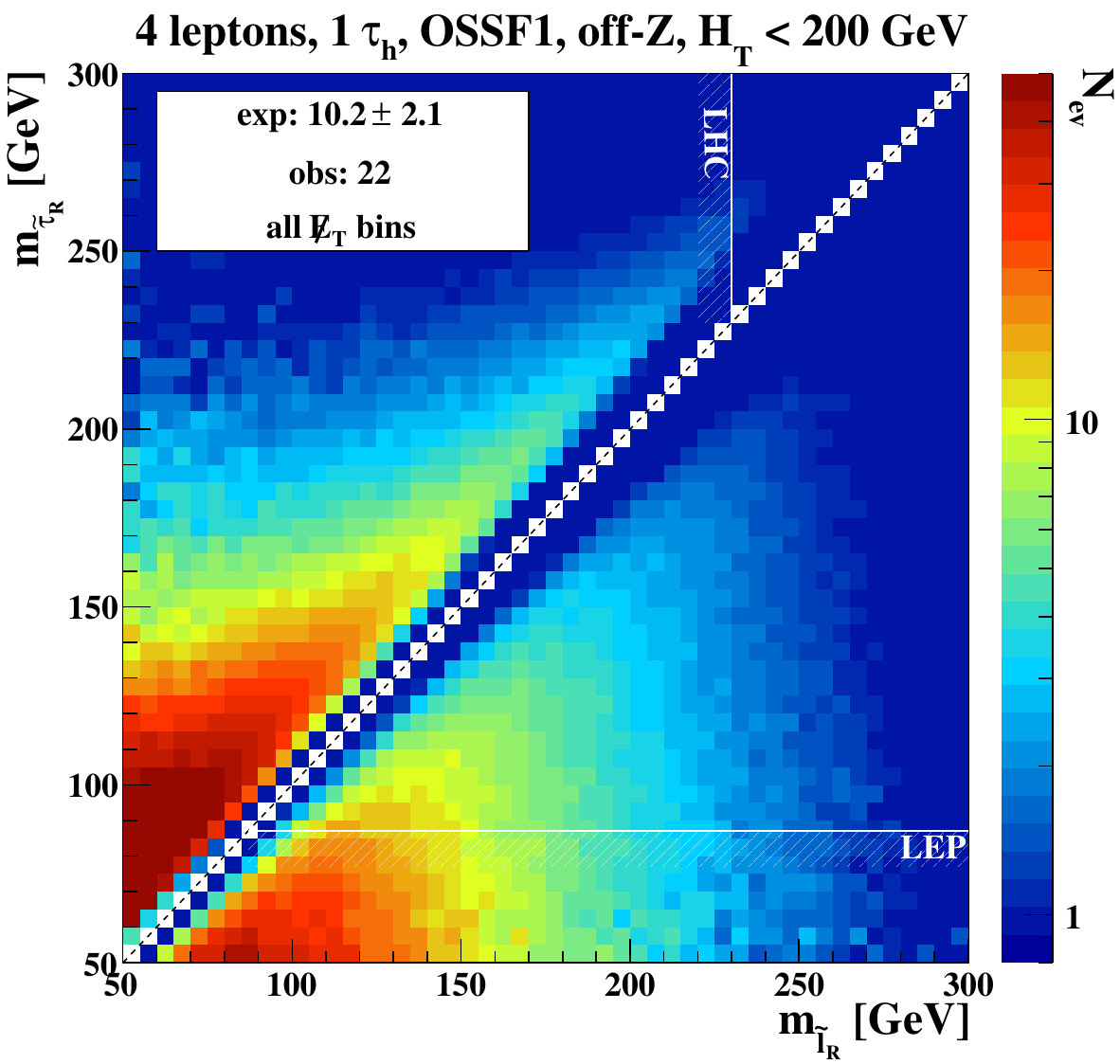}
 \includegraphics[width=.32\textwidth]{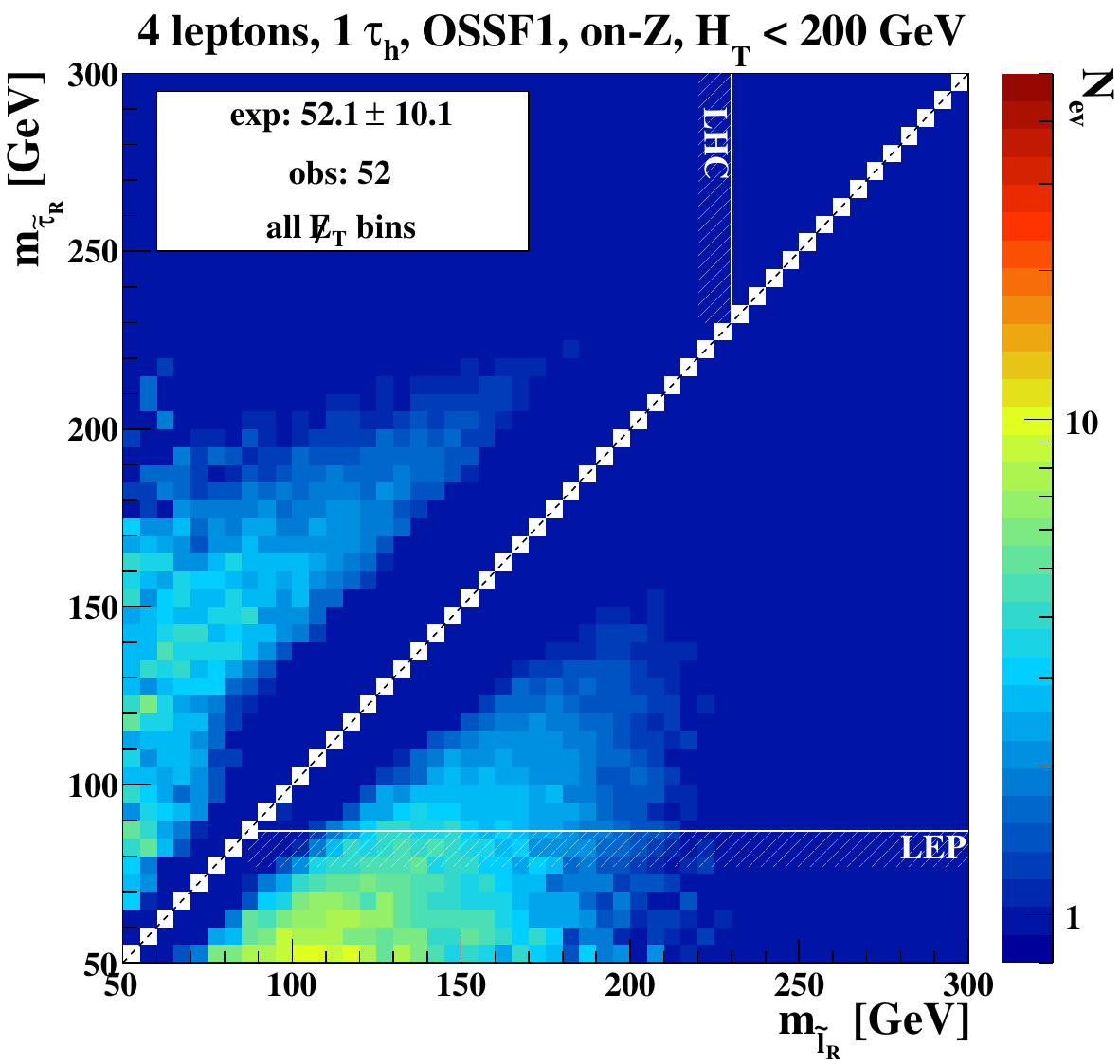}
 \includegraphics[width=.32\textwidth]{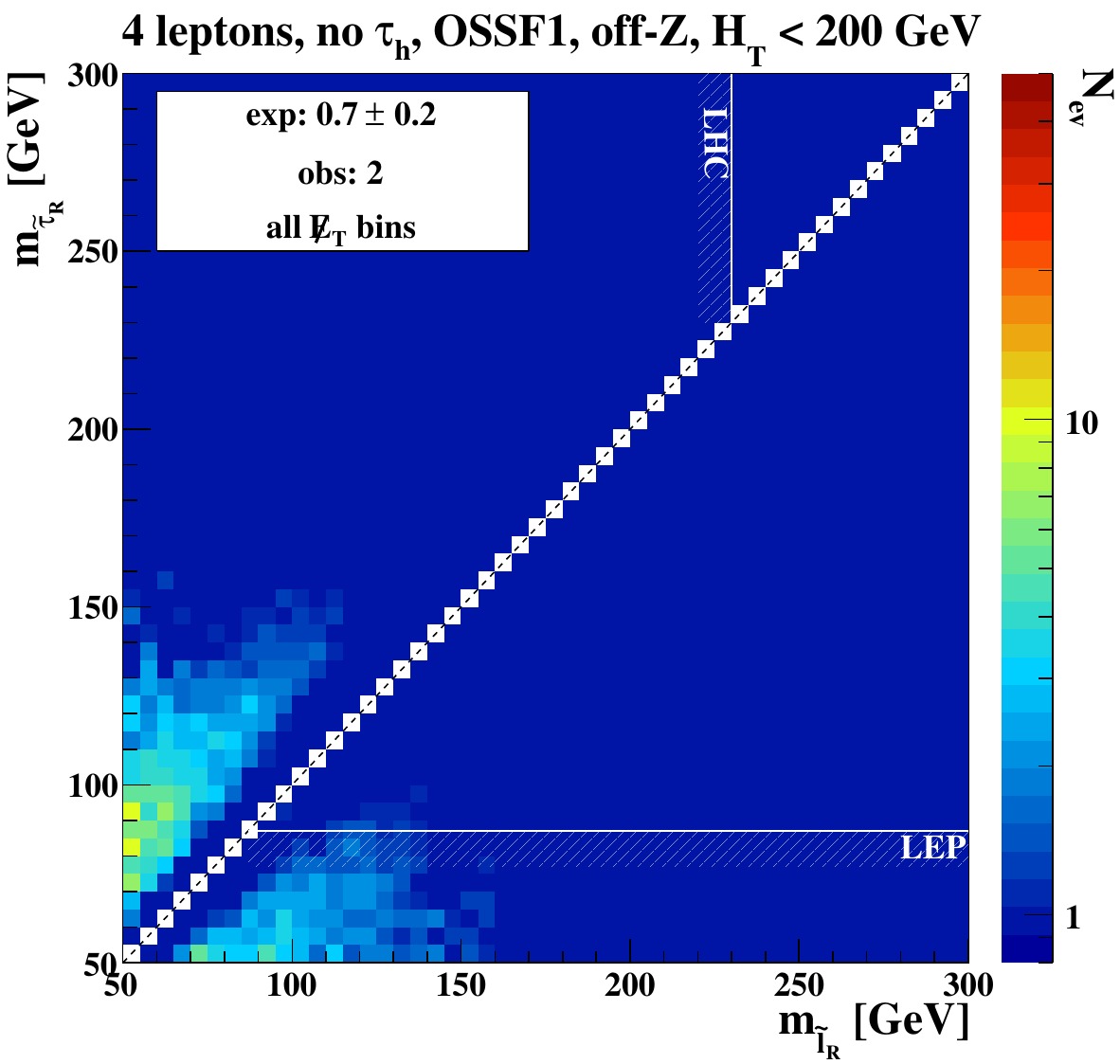}
 \caption{\label{fig:results} The number of signal events in the $m_{\tilde{\ell}_R}/m_{\tilde{\tau}_R}$ mass plane, where we have combined the three $\MET$-bins. The left plot corresponds to the category discussed in the text, where CMS observed an excess. The middle plot corresponds to a similar category, but where the invariant mass of the OSSF lepton pair is instead \emph{within} a window of $\pm15$\,GeV around the $Z$ boson mass. The right plot corresponds to the category where the final state consists of four electrons or muons, but no hadronic tau. In all three plots, the number of
 expected and observed events are indicated, as well as the LHC and LEP exclusion bounds arising from NLSP pair production.} 
\end{figure}

In Figure \ref{fig:results}, we show the number of signal events the processes in Figure \ref{fig:processes} give rise to in the stau/slepton mass plane, where model {\bf M.I}/{\bf M.II} corresponds to the lower/upper triangular half plane. Figure \ref{fig:results} (left) corresponds to the final state category where CMS observed the excess, and we see that both model {\bf M.I} and {\bf M.II} contain regions in the mass plane where the number of signal events fill the gap between the observed and expected number of events. 

So far we have only discussed the multilepton final states arising from the pair production of the NNLSP. Of course, in these models, there will also be pair production of the NLSP. In model {\bf M.II}, the pair produced NLSP sleptons decay to the final state $\ell^+\ell^- {+}\MET$. The current bound on such right-handed sleptons is $m_{\tilde{\ell}_R}>245$ GeV \cite{ATLASsleptons}, which actually excludes the entire parameter space region of {\bf M.II} relevant for explaining the CMS excess. 

In model {\bf M.I}, the pair produced NLSP staus decay to the final state $\tau^+\tau^-{+}\MET$. The current strongest bound on such right-handed staus is still the one set by LEP at \mbox{$m_{\tilde{\tau}_R}>87$ GeV \cite{LEPstau}.} As can be seen from Figure \ref{fig:results}, even if this stau mass bound imposes a non-trivial constraint on {\bf M.I}, there still remains a parameter space region of {\bf M.I} that can explain the CMS excess. The best fit model, obtained by considering the three $\MET$-bins individually, has $m_{\tilde{\ell}_R}=145$\,GeV and $m_{\tilde{\tau}_R}=90$\,GeV. 

In Figure \ref{fig:results} (middle) and (right), we show the number of signal events we obtain for two of the other final state categories in \cite{CMSmultileptons}. We see that the best fit model gives rise to a  number of signal events that is within the 1$\sigma$ variation from the SM prediction in the middle plot. Also in the right plot there is no conflict with the data. We refer the reader to the paper \cite{multileptons} for discussions about data from other categories and searches.   

Concerning the possibility to probe this best fit model, it is important to realize that, in the last decay step of the process in Figure \ref{fig:processes} (left), each of the 90 GeV staus decays to two approximately massless particles, a gravitino and a tau, which therefore roughly share the energy. Since the taus dominantly decay hadronically, many of these signal processes give rise to at least two hadronic taus which are hard enough to allow for reconstruction. We find that the most promising search channel for the best fit model would be in the final state involving two hadronic taus, two or three electrons or muons and $\MET$. In paper \cite{multileptons} we estimate the number of signal events the best fit model would contribute with to these final states, and we find that, already with the existing data set, such a search could have very good sensitivity.  

\section{Multiphoton signatures}
\label{multiphotons}

If SUSY is realized in Nature it must be in a broken phase at low energies. A model-independent consequence of (global) SUSY breaking is the presence of a spin 1/2 Goldstone mode, the goldstino. Upon coupling to gravity, the goldstino is eaten by the spin 3/2 gravitino, becoming its longitudinal components, and the gravitino becomes massive. This is in analogy with electroweak symmetry breaking in the SM, where the Goldstone bosons are eaten by the $Z$ and $W$ bosons, which become massive. When the mass of the gravitino is small compared to the energy scale under consideration, in analogy with the equivalence theorem in the SM, the gravitino can be replaced by its longitudinal goldstino components. We consider this case, where the gravitino mass is small and where the communication of SUSY breaking to the visible sector, which we take to be the MSSM, is done via gauge interactions (i.e.~within the framework of GMSB). 

In this section we investigate how the usual phenomenology of GMSB models is modified if we make the non-minimal assumption that SUSY is broken in more than one hidden sector.\footnote{Also for such a multiple breaking of a symmetry, there is a SM analogy since the electroweak symmetry is broken both by the vacuum expectation value of the SM scalar field and by the chiral condensate in QCD.} If SUSY is broken in  $n$ hidden sectors, there will be $n$ neutral spin 1/2 goldstino-like fermions in the  spectrum. However, there is only one particular linear combination that corresponds to the true goldstino mode, i.e.~the one that is eaten by the gravitino. The remaining $n{-}1$ linear combinations  correspond to so called pseudo-goldstini which, in contrast to the true goldstino, are not protected by the Goldstone shift symmetry and therefore they  acquire masses, both at the tree and the radiative level. In comparison to standard GMSB models, where it is assumed that SUSY is broken in only one hidden sector and that there is only the nearly massless gravitino below the SM superpartners, multiple hidden sector models will, in addition, contain a tower of these massive pseudo-goldstini.  

In the case where the lightest SM superpartner is a Bino-like neutralino, its dominant decay channel is generically to a photon and the heaviest of the pseudo-goldstini. The reason is that the strength of the neutralino couplings are related to the mass of the pseudo-goldstini and the larger the mass, the stronger the coupling. In paper \cite{multiphotons} we also show that, in models with more than two hidden sectors, i.e.~with more than one pseudo-goldstino, the heaviest pseudo-goldstino can decay promptly to a photon and a lighter pseudo-goldstino.\footnote{In order for this decay to be prompt, the SUSY breaking scales should be hierarchical with at least two of them having values around or below $5{-}10$ TeV. See \cite{lowscale} for recent discussions concerning low scale SUSY breaking and \cite{gaugegrav} for some discussions on how it could be realized using the gauge/gravity correspondence.} 
 Hence, in comparison to standard GMSB, the final states will contain additional photons. 

In order to illustrate the characteristic features of GMSB models with multiple hidden sectors and with a Bino-like neutralino being the lightest SM superpartner, let us for concreteness take the number of hidden sectors to be three, and thereby the number of massive pseudo-goldstini to be two. Above the Bino-like neutralino, depending on the ultraviolet model one has in mind, one could consider different superpartners which could be produced at the LHC. Here, as an example, we include the right-handed sleptons in the simplified model we consider. For simplicity, we take all three families of sleptons to be mass-degenerate, and with a change of notation with respect to the Section \ref{multileptons}, ``slepton" here refers to a selectron, smuon or stau, $\tilde{\ell}_R=\tilde{e}_R,\tilde{\mu}_R,\tilde{\tau}_R$. 
\begin{figure}
\centering
 \includegraphics[width=.35\textwidth]{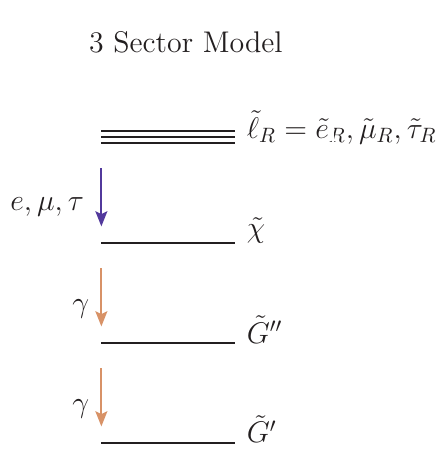}
 \hspace{0.3cm}
 \includegraphics[width=.45\textwidth]{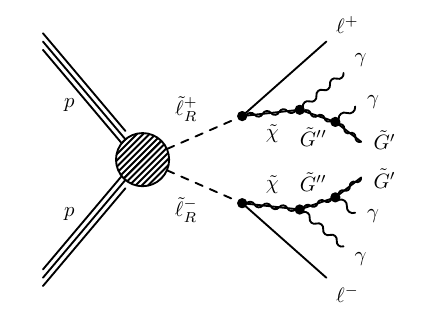}
  \caption{\label{Spec+proc} The spectrum of the simplified model we consider, and the process it gives rise to.}
\end{figure}

In the simplified model depicted in Figure \ref{Spec+proc} (left), the main production mode is slepton pair production and the relevant process is shown in Figure \ref{Spec+proc} (right). In this model, the heaviest pseudo-goldstino, to which the neutralino decays via a photon, is denoted by $\tilde{G}''$. This pseudo-goldstino subsequently decays promptly to the lighter pseudo-goldstino $\tilde{G}'$ which, in contrast to $\tilde{G}''$, is stable on collider time scales. Eventually $\tilde{G}'$ will decay to the gravitino $\tilde{G}$ but since that coupling is very small, this decay takes place outside the detector. Thus, from the point of view of collider physics, the gravitino plays no role and therefore it is not included in Figure \ref{Spec+proc} (left).       

The process in Figure \ref{Spec+proc} (right) gives rise to the final state $\ell^+\ell^- + 4\gamma+\MET$. Note that, if we would consider a different production mechanism by replacing the sleptons by other SM superpartners, then the OSSF lepton pair in Figure \ref{Spec+proc} (right) would be replaced by the corresponding SM partners. In order  to be as model-independent as possible in terms of the production mode, let us focus on the last two decay steps in the process, from which the photons and $\MET$ originate. In comparison to standard GMSB, where each of the two neutralinos decays to a photon and a nearly massless gravitino, since the pseudo-goldstini are massive, the emitted photons will here be softer. Moreover, since some of the $\MET$ that would be carried away by  $\tilde{G}''$ is transformed into soft photon energy in the last decay step, also the amount of $\MET$ will here be smaller.

In paper \cite{multiphotons} we show that, due to the hard cuts on the transverse momentum of the photons ($p_T^\gamma$) and the $\MET$, the existing LHC searches for GMSB are not very sensitive to multiple hidden sector models. Instead, what one should do in order to probe these models is to relax the $p_T^\gamma$ and $\MET$ cuts, but require additional soft photons in the final state. In \cite{multiphotons} we find that a search for four photons, each with $p_T^\gamma>20$\,GeV, and $\MET>$50\,GeV could easily lead to a discovery (or very strong constraints) already with the existing amount data, i.e.~\mbox{20 fb${}^{-1}$ at $\sqrt{s}=8$\,TeV.} 

\section{Conclusions}

In the first part of this note we provided a possible explanation of a slight excess, observed by the CMS collaboration, in terms of a simplified model of GMSB, with a gravitino LSP, a stau NLSP at around 90\,GeV and mass-degenerate selectron/smuon at around 145\,GeV. Since the stau mass is close to the LEP bound it will be interesting to see if the LHC at some point will be able to set a stronger bound on the stau mass. The search we propose that would best probe this model is in the final state $2\tau_h+(2/3)\ell+\MET$. 

In the second part, we discussed GMSB models in which SUSY is broken in more than one hidden sector. The two key features of such multiple hidden sector models were that the final state spectrum was generically softer than in standard GMSB, making the existing LHC searches poorly sensitive, and the presence of additional decay steps, where soft photons are emitted, opening up new search channels. The search we propose, that would be model-independent in terms of the production mode, is an inclusive search in the final state $(3/4)\gamma+\MET$, with minimal $p_T$-requirements on the photons. We focused on the example of slepton pair production which, due to the presence of a lepton pair in the final state, could be probed by a search in the final state $2\ell +2\gamma+\MET$, which would capture also the case where one or two photons are too soft to allow for reconstruction.      

\section*{Acknowledgments}

I would like to thank J.\,D'Hondt, K.\,De Causmaecker, G.\,Ferretti, B.\,Fuks, A.\,Mariotti, K.\,Mawatari and D.\,Redigolo for the collaborations on the projects I presented here. This work is supported by the Swedish Research Council (VR) under the contract 637-2013-475, by IISN-Belgium (conventions 4.4511.06, 4.4505.86 and 4.4514.08) and by the ``Communaut\'e Fran\c{c}aise de Belgique" through the ARC program and by a ``Mandat d'Impulsion Scientifique" of the F.R.S.-FNRS. 
Finally I would like to thank the organizers of the conference ``Rencontres de Moriond 2014, Electroweak Session" for their effort in organizing a very nice and fruitful meeting.

\end{document}